\documentclass[reprint]{aastex62}

\accepted{for publication in The Astrophysical Journal}

\shorttitle{Transmission and Emission of hot-Jupiters}
\shortauthors{Chakrabarty and Sengupta }
\usepackage{natbib}
\begin{document}

\title{EFFECTS OF THERMAL EMISSION ON THE TRANSMISSION SPECTRA OF HOT JUPITERS }

\correspondingauthor{Aritra Chakrabarty}
\email{aritra@iiap.res.in}

\author[0000-0001-6703-0798]{Aritra Chakrabarty}
\affil{Indian Institute of Astrophysics, Koramangala 2nd Block, Bangalore 560034, India}
\affil{University of Calcutta, Salt Lake City, JD-2 Kolkata 750098, India}

\author[0000-0002-6176-3816]{Sujan Sengupta}
\affil{Indian Institute of Astrophysics, 
Koramangala 2nd Block, Bangalore 560034, India}

\begin{abstract}
The atmosphere on the dayside of a highly irradiated close-in gas giant (also known as a hot Jupiter) absorbs a significant part of the incident stellar radiation which 
again gets re-emitted  in the infrared wavelengths both from the day and the night sides of the planet. The re-emitted thermal radiation from the night side facing
the observers during the transit event of such a planet contributes to the transmitted stellar 
radiation. We demonstrate that the transit spectra at the infrared region get altered significantly when such re-emitted
thermal radiation of the planet is included. We assess the effects of the 
thermal emission of the hot Jupiters on the transit spectra by simulating 
observational spectroscopic data with corresponding errors from the different channels of the upcoming James Webb Space Telescope. 
We find that the effect is statistically significant with respect
to the noise levels of those simulated data. Hence, we convey the important 
message that the planetary thermal re-emission must be taken into consideration in the retrieval models of transit spectra for hot Jupiters
 for a more accurate interpretation of the observed transit spectra.           

\end{abstract}


\section{Introduction} \label{sec:intro}

Transit spectroscopy is an essential tool for probing into the upper 
atmospheres of the close-in exoplanets \citep{charbonneau02,sing16,sengupta20}. As an exoplanet transits across its parent star, a fraction of the starlight is transmitted through the planet's upper atmosphere. Light at specific wavelengths is preferentially absorbed, depending on the atmospheric chemical composition and physical properties. An accurate interpretation of the observed transit spectra, however, requires a self-consistent theoretical model that must incorporate the physical and chemical properties of the planetary atmospheres in sufficient detail, including scattering albedo  \citep[e.g.][]{dekok12,sengupta20} and thermal re-emission from the night side (this work). Theoretical models for the planetary transit spectra presented by various groups are described in articles by \cite{brown01, tinetti07, madhu09, burrows10, fortney10, line12, waldmann15b, line15, kempton17, goyal19}; etc. All of these models have only considered the effect of the total extinction coefficient and ignored the effect of scattering albedo in calculating the transmission depth at different wavelengths. Recently \cite{sengupta20} demonstrated that the diffused reflection and transmission due to scattering of the transmitted starlight by the atoms and molecules affect the transmitted flux and hence, the transit depth significantly in the optical. Therefore, modeling the transit spectra correctly and consistently requires one to solve the multiple scattering radiative transfer equations. However, \cite{sengupta20} show that effect of scattering albedo on transmission spectra is less than 10 ppm in the infrared region.

On the other hand, because of the extreme proximity to their parent stars, many gas giants are extremely hot. As they are tidally locked with their host stars, the dayside of such a planet is so hot due to the intense irradiation that its equilibrium temperature for zero albedo, $T_{eq0}$ can reach as high as $\sim$4000 K \citep{gaudi17}. 
The heat is redistributed to the night side by the advection process and the night side of the planet facing the observer during transit also becomes hot depending on the heat re-circulation efficiency, $\epsilon$ of the planetary atmosphere. The average temperature $T_n$ of the night side of the planet can be estimated by the relation \citep{cowan11, keating17}
\begin{equation}\label{eq:tn}
 T_n = T_{eq0} (1-A_B)^\frac{1}{4}\epsilon^\frac{1}{4} = T_{n0}(1-A_B)^\frac{1}{4}
\end{equation}
 where $A_B$ is the Bond albedo of the atmosphere. We define $T_{n0}$ as the night-side average temperature for zero Bond albedo. Even with a small value of $\epsilon$, the average night side temperature $T_n$ can be quite high if $T_{eq0}$ is high enough.  Some previous studies provide an estimation of the night-side temperature of a few exoplanets. For example, \cite{keating17} report $T_n = 1080 \pm 11$ K  for WASP-43 b; \cite{demory16} report a night-side brightness temperature of $1380 \pm400$ K for 55 Cancri e based on their observation in the 4.5-$\mu m$ channel of the Spitzer Space Telescope Infrared Array Camera (IRAC); \cite{arcangeli19} report $T_n \leq 1430$ K  for WASP-18 b at  3-$\sigma$ level; etc. Such a hot region facing the observer would emit radiation in the infrared wavelengths which should be added up with the transmitted stellar radiation. Thus the transmitted flux would be affected by the re-emission of hot planets in the longer wavelengths.     

In this paper, we demonstrate the effect of thermal re-emission on the transit spectra of the hot Jupiters of different size and surface gravity and with equilibrium temperature ranging from 1200K to 2400K.
In Section \ref{sec:trandep} we provide the formalisms for calculating the transit depth with and without thermal re-emission. Section \ref{sec:calctransem} outlines the detailed procedure followed to calculate the transmitted and the re-emitted flux from the hot-Jupiters. In Section \ref{sec:ptoa} we discuss the 1D pressure-temperature grids calculated and the databases adopted for the calculation of abundance and absorption and scattering opacity for the modeling of the atmospheres of the hot-Jupiters. In Section \ref{sec:cssd} we discuss the results from all the case studies as well as the results from the testing of detectability of the effect of thermal emission on the transit spectra of hot Jupiters by simulating observational transit spectroscopic data to be observed from the upcoming James Webb Space Telescope (JWST) using the open-source Pandexo code \citep{batalha17} available in the public domain \footnote{\url{https://github.com/natashabatalha/PandExo}}. In the last section, we conclude the key points.

\section{The Transit Depth with Planetary Thermal Re-emission} \label{sec:trandep}

The transit spectra of an exoplanet are expressed in term of transit depth 
$D(\lambda)$ which is the difference in stellar flux during out of transit
and during the transit of the planet and normalized to the unblocked stellar
flux. When the planetary radiation does not contribute to the stellar flux, 
it can be written as \citep[e.g.][]{kempton17, sengupta20}

\begin{equation}\label{eq:tdne1}
D_{NE}(\lambda) = 1 - \frac{F_{in}}{F_{out}}=1-\frac{(1-\frac{R_{PA}^2}{R_*^2})
F_* + F_P}{F_*}
\end{equation}
where, $D_{NE}$ is the transit depth with no thermal radiation from  the 
transitting planet, $R_{PA}$ is the sum of the base radius $R_P$ and the 
atmospheric height of the planet, $F_{in}$ and $F_{out}$ are the in-transit and 
out-of-transit stellar flux respectively. In case of pure transmission when
thermal emission of the planet is ignored, $F_{out}$ is the stellar flux $F_*$.
$F_P$ is the portion of the stellar flux which is transmitted through the upper atmosphere of the planet and undergoes absorption and
scattering through the medium. The above equation can also
be written as:    
\begin{equation}\label{eq:tdne2}
D_{NE}(\lambda) = \frac{R_{PA}^2}{R_*^2} - \frac{F_P}{F_*}
\end{equation}

For a transiting planet with a hot night side facing the observer, the 
re-emitted radiation flux $F_{Th}$ is added to the observed fluxes.
Hence, the transit spectra including the effect of the
re-emission from the planet can be expressed as:

\begin{equation}\label{eq:tde}
D_E(\lambda) = 1-\frac{(1-\frac{R_{PA}^2}{R_*^2})
F_* +F_{Th}\frac{R^2_{PA}}{R_*^2}+F_P}{F_*+F_{Th}\frac{R^2_{PA}}{R^2_*}}
\end{equation}

A transit spectrum is usually produced from observation by calculating the transit depth at different wavelengths normalized with respect to the baseline flux, i.e., the flux observed right before ingress or immediately after egress. One of the methods is to calculate the wavelength-dependent transit depth from the light curves at different wavelength bins extracted from the time-series spectra of the host stars observed during a transit event using space-based instruments like HST+STIS, HST+WFC3, etc. \citep[][etc.]{charbonneau02,gibson12,berta12,deming13,sing16} or ground-based instruments like VLT+FORS, VLT+FORS2, Magellan+MMIRS, GEMINI-N+GMOS, GTC+OSIRIS etc \citep[][etc.]{bean11, stevenson14, nikolov16, palle16, huitson17}. The wavelength-dependent transit depth can also be calculated directly from the photometric light curves at different wavelength bands observed using ground-based instruments or space-based instruments like Spitzer Space Telescope Infrared Array Camera (IRAC) etc. during a transit event \citep{tinetti07,sing16}. In either case, the normalized wavelength-dependent transit depth (equivalently, the wavelength-dependent radius) can be more accurately modeled with the incorporation of the planetary thermal emission $F_{Th}$. However, the impact on the accuracy by omitting the effect of $F_{Th}$ might be, in some cases (e.g., at wavelength shorter than 2 $\mu$m), small compared to the uncertainties in the data. We discuss the significance of the effect of $F_{Th}$ with respect to the uncertainties in the observational data elaborately in Section \ref{sec:cssd}.

Finally, Equation~\ref{eq:tdne1} and Equation~\ref{eq:tde} gives

\begin{equation}\label{eq:tdne2tde}
\frac{D_{NE}(\lambda)}{D_{E}(\lambda)}=1+\frac{F_{Th}}{F_*}
\frac{R_{PA}^2}{R_*^2}.
\end{equation}
 As evident from Equation \ref{eq:tdne2tde} the contribution from 
the thermal re-emission of the planet reduces the transit depth.

\section{Calculations of Transmission and Emission Flux} \label{sec:calctransem}

To calculate the transmitted stellar flux $F_P$ that passes through the
atmosphere of a hot Jupiter, we
 first, calculate the reduced stellar intensity that suffers absorption and
 scattering in the planetary atmosphere and then integrate over the angle
 subtended by the annular region of the atmosphere. \cite{sengupta20} have shown that
 an accurate approach to calculate the reduced intensity is to solve the
 multi-scattering radiative transfer equations that incorporate the diffused
reflection and transmission of radiation due to scattering. 
 The radiative transfer equations including diffused reflection and
 transmission for a plane-parallel geometry can be expressed as
 \citep[e.g.][]{chandrasekhar60,sengupta20}

\begin{equation}\label{eq:radtran}
\mu\frac{dI(\tau_{LOS},\mu,\lambda)}{d\tau_{LOS}}=I(\tau_{LOS},\mu,\lambda)-\frac{\omega}{2}\int_{-1}^1{p(\mu,\mu')I(\tau_{LOS},\mu,\lambda)\text{d}\mu'}-\frac{\omega}{4}F_* e^{-\tau_{LOS}/\mu_0}p(\mu,\mu_0)
\end{equation}

where $I(\tau_{LOS},\mu,\lambda)$ is the specific intensity of the diffused radiation field  along the direction $\mu=\cos\theta$, $\theta$ being the 
angle between the axis of symmetry and the ray path, $F_*$ is the incident 
stellar flux in the direction $-\mu_0$, $\omega$ is the albedo for single 
scattering i.e. the ratio of scattering co-efficient to the extinction 
coefficient, $p(\mu,\mu')$ is the scattering phase function that describes 
the angular distribution of the photon before and after scattering and 
$\tau_{LOS}$ is the optical depth along the line of sight to the observer.
The detail method for calculating $\tau_{LOS}$ is given in \cite{sengupta20}. We adopt the Rayleigh  phase function \citep[e.g.][]{chandrasekhar60} for cloud-free atmospheres. For cloudy atmospheres, we use the Mie phase function. This is elaborated in Section \ref{subsec:clouds}.
The planetary thermal emission is important only at longer wavelengths where the effect of scattering albedo is negligible, less than 10 ppm \citep{sengupta20}, even in the presence of atmospheric clouds. Therefore, in the present work, we have ignored the effect of scattering albedo in the calculation of $F_P$.  

On the other hand, to calculate the flux of the planetary thermal radiation $F_{Th}$, we treat the planet as a self-luminous object and solve the radiative transfer equations in the following form:

\begin{equation}\label{eq:rademn}
\mu\frac{dI(\tau,\mu,\lambda)}{d\tau}=I(\tau,\mu,\lambda)-\frac{\omega}{2}\int_{-1}^1{p(\mu,\mu')I(\tau,\mu,\lambda)\text{d}\mu'}-(1-\omega)B(\tau,\lambda)
\end{equation}
where, $I(\tau,\mu,\lambda)$ denotes the specific intensity of the thermal
 radiation field along the direction $\mu$, $\tau$ is the optical depth in
the radial direction, $B(\tau,\lambda)$ is the Planck function corresponding to the temperature of the atmospheric layer with optical depth $\tau$ at a particular wavelength.

The above radiative transfer equations are solved by using the 
Discrete Space Theory method \citep{peraiah73}.
The numerical method is described in \cite{sengupta20} and in \cite{sengupta09}. $F_P$ and $F_{Th}$
are calculated separately and used into Equation \ref{eq:tdne1} and in
Equation \ref{eq:tde} to derive $D_{NE}$ and $D_E$ respectively.

\begin{figure}[!ht]
\centering
\includegraphics[scale=0.37,angle=0]{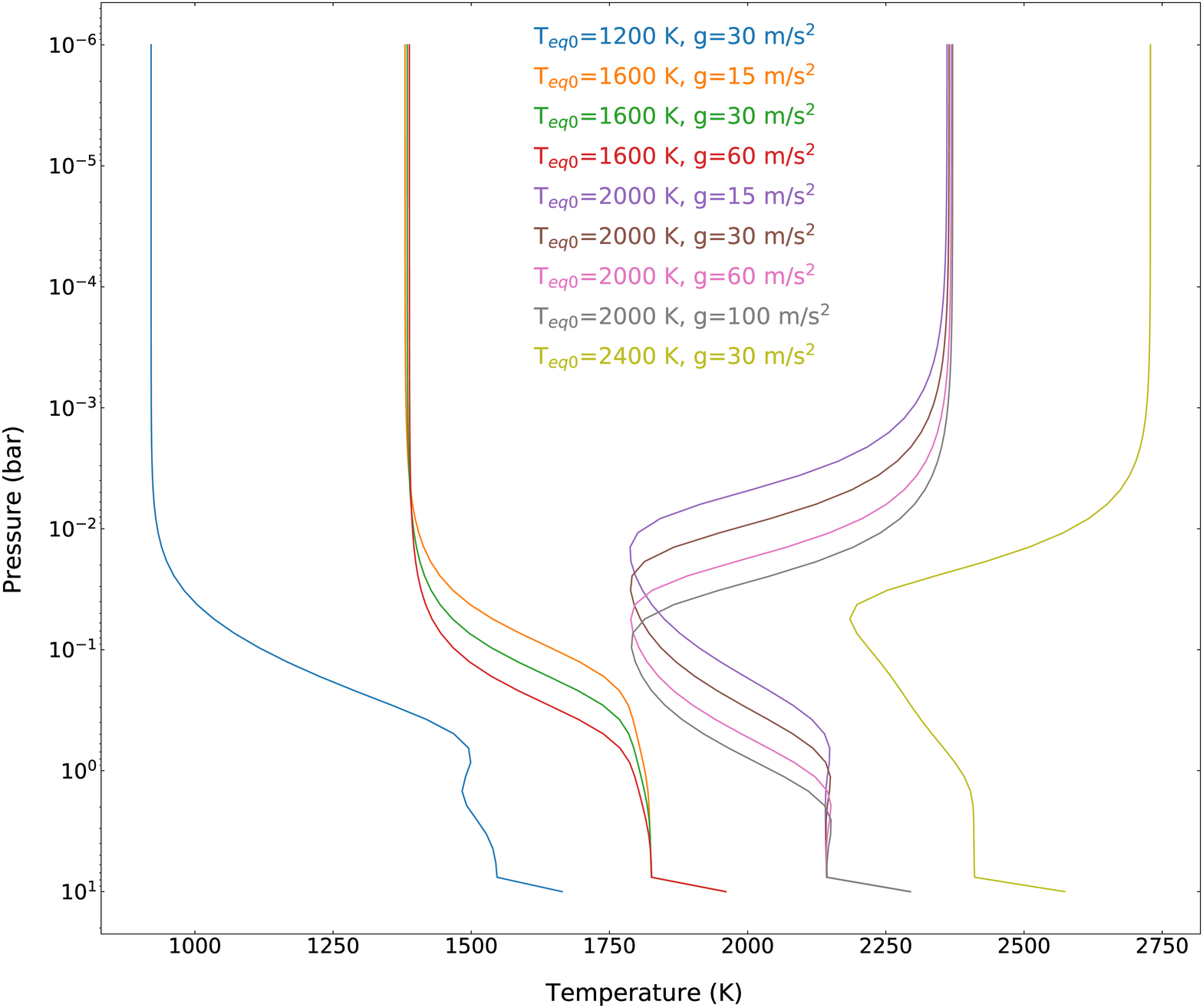}
\caption{Pressure-Temperature profiles for different $T_{eq0}$ and $g$ adopted in this work. For all the cases, the effect of TiO and VO  is included.
\label{fig:ptprof}}
\end{figure}

\section{Pressure-Temperature Grids and the Opacity and Abudance Data} \label{sec:ptoa}

The atmospheric pressure-temperature structure plays an important role not
only in determining the optical depth of the medium but also in estimating
the thermal radiation of the planet. In order to calculate the pressure-temperature profiles, we use the FORTRAN implementation of the 
analytical models of non-Grey irradiated planets provided by 
\cite{parmentier14,parmentier15}. This code is available in public 
domain\footnote{\url{http://cdsarc.u-strasbg.fr/viz-bin/qcat?J/A+A/574/A35}}. It uses the
functional form  for Rosseland opacity derived by \cite{valencia13}. The
Rosseland opacities of \cite{freedman08} are adopted in the derivation.
 For  $T_{eq0}\geq 1700 K$, 
the pressure-temperature (P-T) profiles show temperature inversion when
TiO and VO are included. Previous studies, including, e.g., \cite{sengupta20} find that the hot-Jupiters are almost
opaque to the transmitted flux at pressures level below 1 bar. Therefore, in the present work, we have considered
the P-T profiles up to 1 bar pressure level such that the base radii $R_P$ of the hot Jupiters
considered in the calculations of the transmission spectra are located at 1 bar pressure level.

However, the thermal radiation of a hot planet emerges from a deeper layer of
the atmosphere and so we have considered the P-T profiles of hot Jupiters
with $T_{eq0}$ ranging between 1200K to 2400K and the surface gravity over
a range of 15-100 m/s$^2$ at 10 bar pressure level.
Figure~\ref{fig:ptprof} shows the P-T profiles for all the 
case-studies with different $T_{eq0}$.

For all the calculations we have adopted solar metallicity and solar 
system abundance for the atoms and the molecules present in the atmospheres.
We have considered 28 molecular and atomic species as mentioned in \cite{sengupta20}.
The abundance for all these atoms and molecules has been calculated using
the abundance database given in the open-source package 
``Exo-Transmit" \citep{kempton17} available in the public domain \footnote{\url{https://github.com/elizakempton/Exo_Transmit}}. In this package, the abundances for major
atmospheric constituents as a function of temperature and pressure are calculated based on the solar system abundances of \cite{lodders03}. Also, to calculate the absorption and 
scattering coefficients we have used the opacity database from the same 
package. These opacities are based on the molecular databases of \cite{freedman08} and \cite{freedman14}. The line lists used to generate the molecular opacities are also tabulated by \cite{lupu14}. Both the abundance and opacity databases are available over a broad grid of pressure and temperature from which we have interpolated to our particular P-T profiles. We have adopted the equation of state (EOS) for rain-out condensation. We have also calculated the opacity due to atmospheric clouds comprising mainly of amorphous Forsterite by using Mie theory (see
 Section \ref{subsec:clouds}).
 
Modeling an atmosphere with a high day-night temperature contrast is not straightforward because  
1D pressure-temperature (P-T) profile may not be adequate in such a scenario and therefore one requires a 3D P-T mapping using Global Circulation Model (GCM) or a limb-averaged P-T profile \citep{kataria15,kataria16,evans18}.
In order to avoid such complications, we have assumed a hot-Jupiter atmosphere
 with $\epsilon=1$ ($T_{n0}=T_{eq0}$) i.e. a globally averaged P-T profile in all the cases as our main motive is to demonstrate the effect of the night-side temperature $T_n$. This corresponds to $f=0.25$, where f is the flux parameter as defined in \cite{burrows03} and also, used in \cite{guillot10, parmentier14, parmentier15}; etc.

The night side of the planetary atmosphere may also get warmed up by the release of the internal energy characterized by the internal 
temperature ($T_{int}$) and may cause thermal radiation (self-emission) but this radiation should be insignificant as
compared to the thermal re-emission from planet older than $100$ Myr \citep{burrows97} in age.

\section{Case Studies: Simulation and Testing of Detectability} \label{sec:cssd}

We checked the detectability of the effect of thermal emission on the transit spectra by using the simulated data. We also investigate the extent to which the transit spectra $D_E$ depend on the planetary properties such as $T_n$, $R_P/R_*$, the atmospheric clouds and the surface gravity as well as the spectral types of the host stars. The following subsections describe the results from these case studies.

\begin{figure}[!ht]
\centering
\includegraphics[scale=0.33,angle=0]{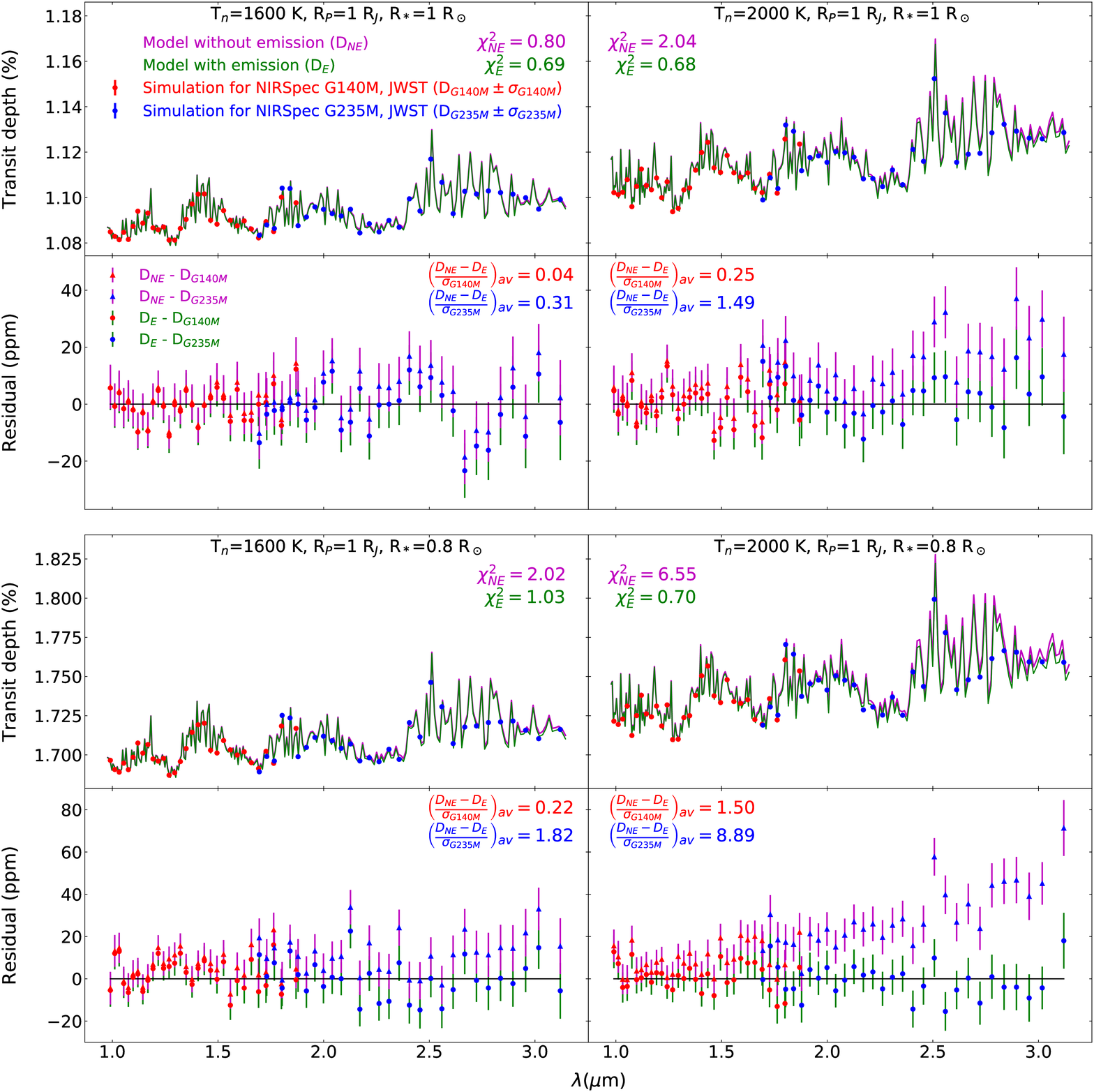}
\caption{Simulated observational data of transit depth with emission combined over 4 observed transit events using instrument modes NIRSpec G140M and NIRSpec G235M of JWST with error-bars, viz. $D_{G140M} \pm \sigma_{G140M}$ (red) and $D_{G235M} \pm \sigma_{G235M}$ (blue) respectively, are shown in the top panels assuming a G2V host star with J-band magnitude = 8. Models without and with thermal emission ($D_{NE}$ in magenta and $D_E$ in green respectively) for different values of $T_n$ and $R_*$ with $R_P$ = 1 $R_J$ are shown with corresponding chi-square values, keeping g fixed at g = 30 m/s$^2$. The bottom panels show the difference between the model without emission and the simulated observational data (red and blue triangles with magenta error-bars) as compared to the  difference between the the model with emission and the same simulated observational data (red and blue circles with green error-bars). Also, the mean of the ratio of the difference between the two models to the noise levels of the two modes, viz. $\left(\frac{D_{NE}-D_E}{\sigma_{G140M}}\right)_{av}$ and $\left(\frac{D_{NE}-D_E}{\sigma_{G235M}}\right)_{av}$, are shown in the bottom panels.
\label{fig:jwst1}}
\end{figure}

\begin{figure}[!ht]
\centering
\includegraphics[scale=0.33,angle=0]{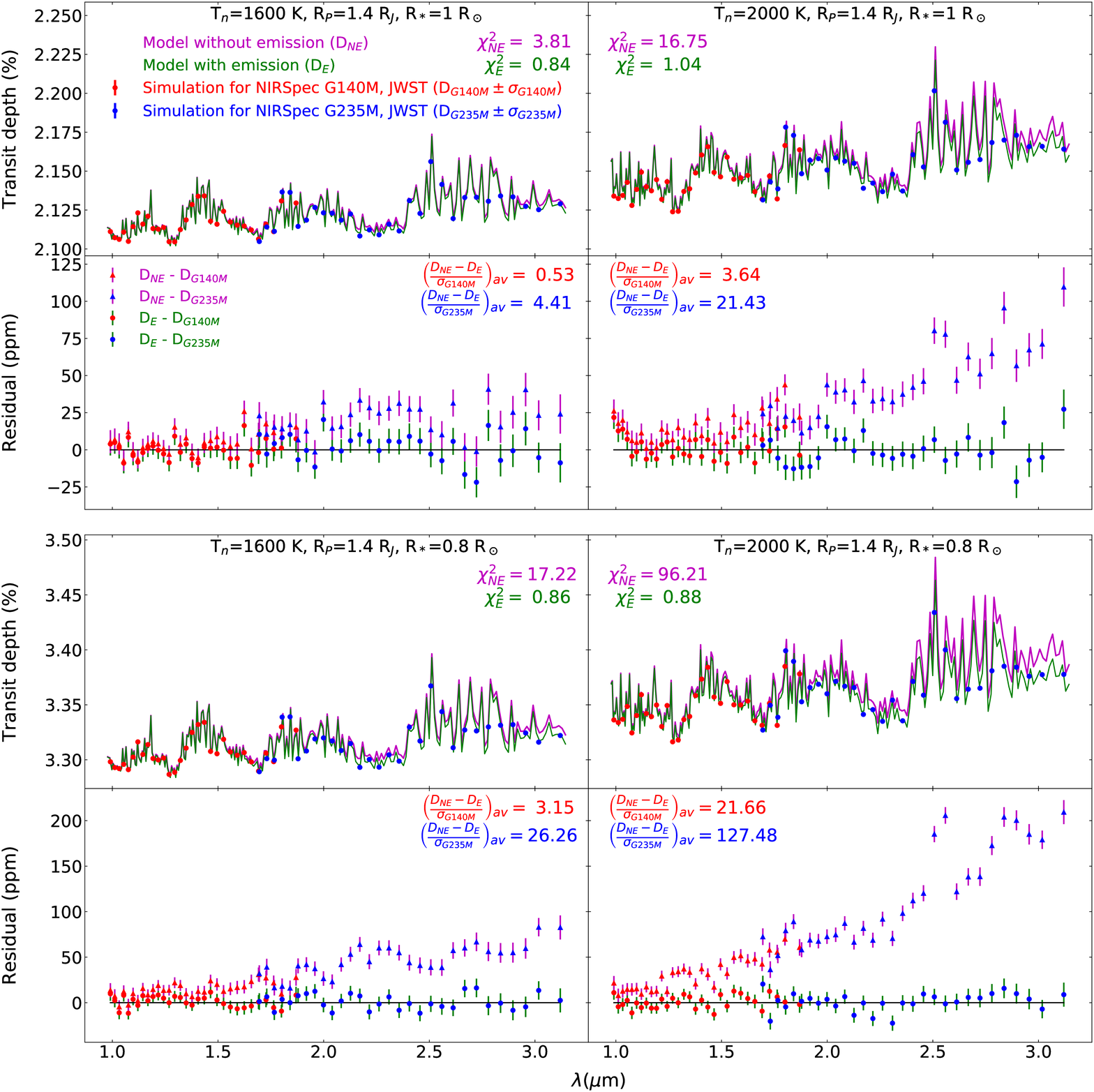}
\caption{Same as Figure \ref{fig:jwst1} but with $R_P$ = 1.4 $R_J$.
\label{fig:jwst2}}
\end{figure}

\begin{figure}[!ht]
\centering
\includegraphics[scale=0.33,angle=0]{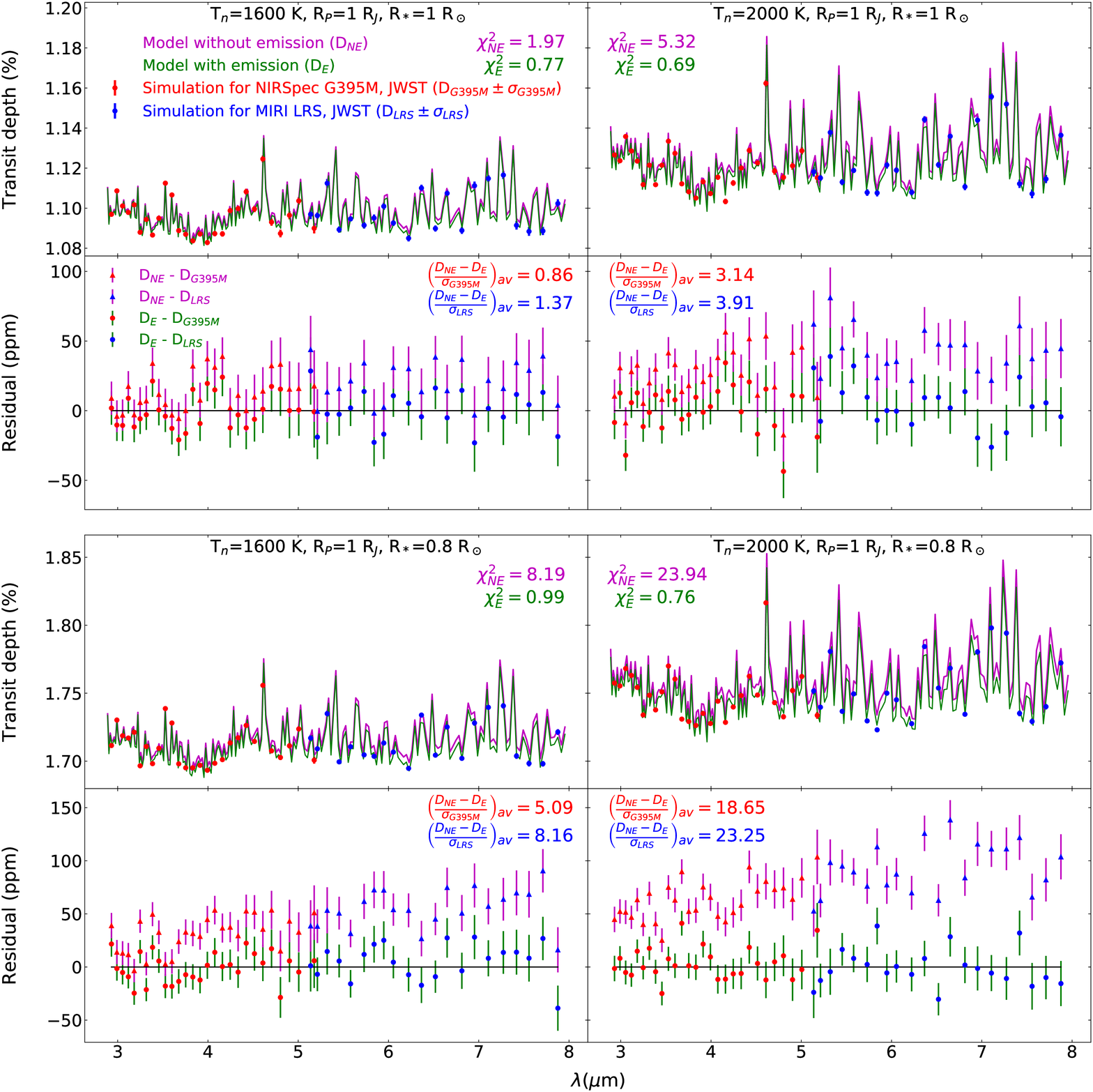}
\caption{Simulated observational data of transit depth with emission combined over 4 observed transit events using instrument modes NIRSpec G395M and MIRI LRS (slitelss) of JWST with error-bars, viz. $D_{G395M} \pm \sigma_{G395M}$ (red) and $D_{LRS} \pm \sigma_{LRS}$ (blue) respectively, are shown in the top panels assuming a G2V host star with J-band magnitude = 8. Models without and with thermal emission ($D_{NE}$ in magenta and $D_E$ in green respectively) for different values of $T_n$ and $R_*$ with $R_P$ = 1 $R_J$ are shown with corresponding chi-square values, keeping g fixed at g=30 m/s$^2$. The bottom panels show the difference between the model without emission and the simulated observational data (red and blue triangles with magenta error-bars) as compared to the  difference between the the model with emission and the same simulated observational data (red and blue circles with green error-bars). Also, the mean of the ratio of the difference between the two models to the noise levels of the two modes, viz. $\left(\frac{D_{NE}-D_E}{\sigma_{G395M}}\right)_{av}$ and $\left(\frac{D_{NE}-D_E}{\sigma_{LRS}}\right)_{av}$, are shown in the bottom panels.
\label{fig:jwst3}}
\end{figure}
 
\begin{figure}[!ht]
\centering
\includegraphics[scale=0.33,angle=0]{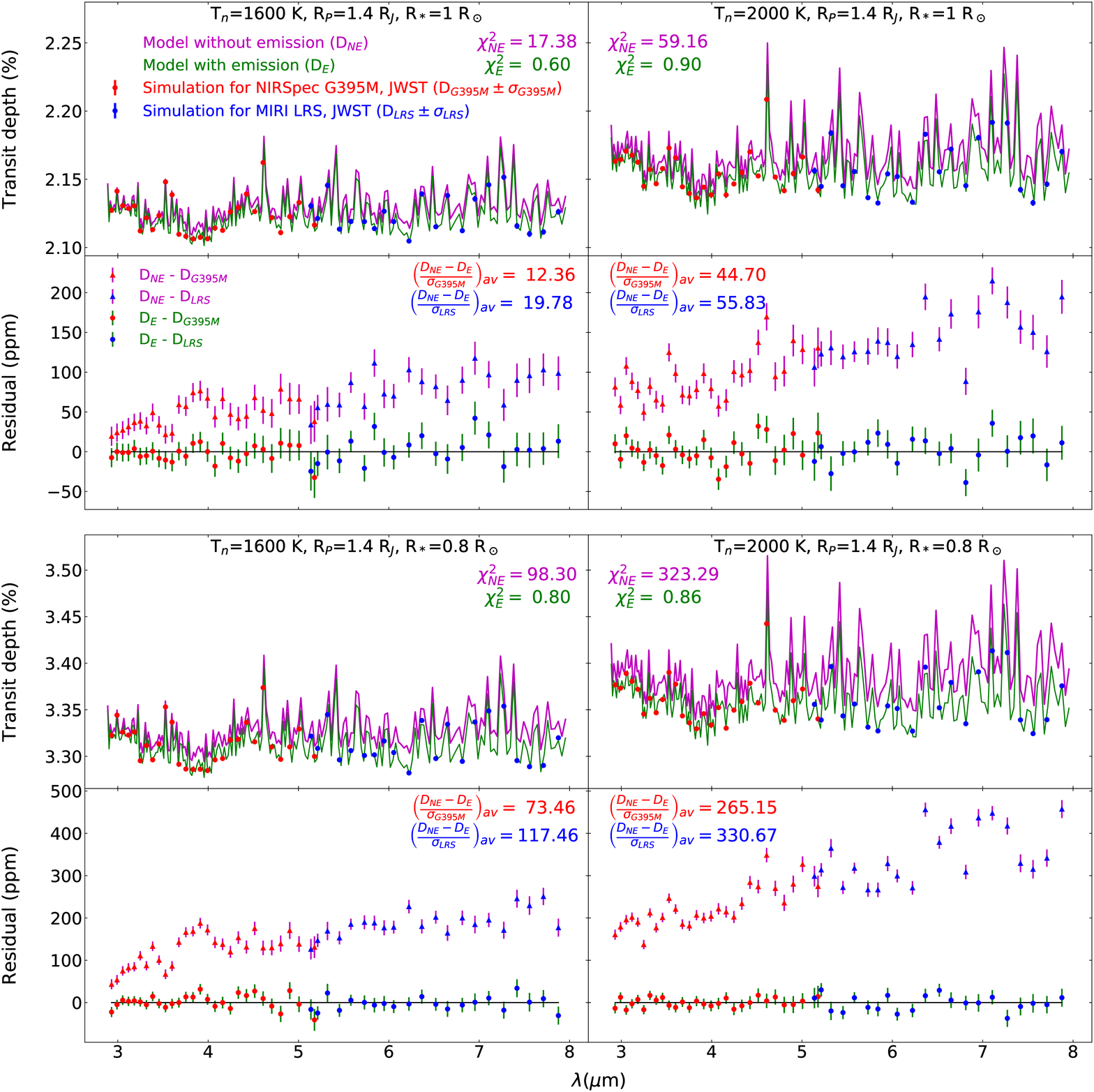}
\caption{Same as Figure \ref{fig:jwst3} but with $R_P$ = 1.4 $R_J$.
\label{fig:jwst4}}
\end{figure} 

\subsection{Detectability with JWST}

To understand the significance of the effect of  thermal emission from 
transiting hot Jupiters on the transit spectra, we need to compare the
difference between $D_{NE}$ and $D_E$ with the noise levels of the actual
observed data. For that purpose, we have focused on the observing capability
 of the upcoming James Webb Space Telescope (JWST) as this mission is going
 to be at the forefront of exoplanet characterization. Considering
the contribution of thermal emission from hot Jupiters to the transit spectra, we use our model calculations for $D_E$ and,
simulate some observational transit spectra with error-bars that can be
 observed using the IR instruments of JWST. The simulation is done by
 using the open-source code Pandexo \citep{batalha17}. 
A host star of spectral type G2V with J-mag=8 and a saturation level equal to 
70 $\%$ of the full well potential are considered for this simulation. The spectra with noise-levels with a resolution of $R \sim 50$ are calculated by combining the data over 4 observed transit events, each with a duration of  2 hours ($T_{14}$). We have considered the instrument modes viz. NIRSpec G140M and NIRSpec G235M for the wavelength region 1-3 $\mu m$ and NIRSpec G395M and MIRI LRS (slitless) for the wavelength region 3-8 $\mu m$ (see Table 1 of \cite{batalha17}) and the corresponding simulated transit depth and noise levels are denoted by $D_{G140M} \pm \sigma_{G140M}$, $D_{G235M} \pm \sigma_{G235M}$, $D_{395M} \pm \sigma_{G395M}$ and $D_{LRS} \pm \sigma_{LRS}$ respectively. For each pair of instrument modes we present the comparison of the models ($D_{NE}$ vs $D_E$), constructed using different sets of planetary paramters, with the simulated data along with their residuals in Figure \ref{fig:jwst1}-\ref{fig:jwst4}.
 In Figure \ref{fig:jwst1} and Figure \ref{fig:jwst3} the model parameters are (i) $T_n=1600$K, $R_P=1$ $R_J$, $R_*=1$ $R_\odot$, (ii) $T_n=2000$ K, $R_P=1$ $R_J$, $R_*=1$ $R_\odot$, (iii) $T_n=1600$ K, $R_P=1$ $R_J$, $R_*=0.8$ $R_\odot$, and (iv) $T_n=2000$ K, $R_P=1R_J$, $R_*=0.8R_\odot$. In Figure \ref{fig:jwst2} and Figure \ref{fig:jwst4}, the model parameters are (i) $T_n=1600$ K, $R_P=1.4$ $R_J$, $R_*=1.4$ $R_\odot$, (ii) $T_n=2000$ K, $R_P=1.4$ $R_J$, $R_*=1$ $R_\odot$, (iii) $T_n=1600$ K, $R_P=1.4$ $R_J$, $R_*=0.8$ $R_\odot$, and (iv) $T_n=2000$ K, $R_P=1.4R_J$, $R_*=0.8R_\odot$. The figures also show the chi-square values of the models (top sub-panels) and the mean of the ratio of ($D_{NE}-D_E$) to the 1-$\sigma$ noise-levels of the above modes (viz. $\sigma_{G140M}$, $\sigma_{G235M}$, $\sigma_{G395M}$, and $\sigma_{LRS}$).

These figures show that the difference between $D_{NE}$ and $D_E$ for $R_P/R_*=1$ is of no or extremely low significance ($<20$ ppm and $<4$-$\sigma$). Also, at wavelengths up to 2 $\mu$m, the difference between $D_{NE}$ and $D_E$ is significant ($>25$ ppm and $\sim 22$-$\sigma$) only for $T_n=2000$ K, $R_P=1.4R_J$, $R_*=0.8R_\odot$. For all other combinations of $T_n$, $R_P$, and $R_*$, the difference is of no or low significance ($<25$ ppm and $<4$-$\sigma$).  However, for wavelengths longer than 2 $\mu$m, we find that the difference between the models increases with increasing $T_n$ and $R_P/R_*$ and reaches up to 500 ppm (330-$\sigma$) for $T_n=2000$ K, $R_P=1.4R_J$, $R_*=0.8R_\odot$ . The chi-square values shown in these figures imply that for higher values of $T_n$ and $R_P/R_*$ and wavelength $\gtrsim2$ $\mu m$, the simulated data are fitted well with the model transit spectra only when planetary thermal emission is incorporated.  This demonstrates the fact that in order to achieve the precision  level of the instruments on-board JWST, the effect of thermal 
emission from  hot Jupiters must be taken into consideration in the retrieval model for transit spectra.

\subsection{Host Stars of Different Spectral Types}

\begin{figure}[ht]
\centering
\includegraphics[scale=0.4,angle=0]{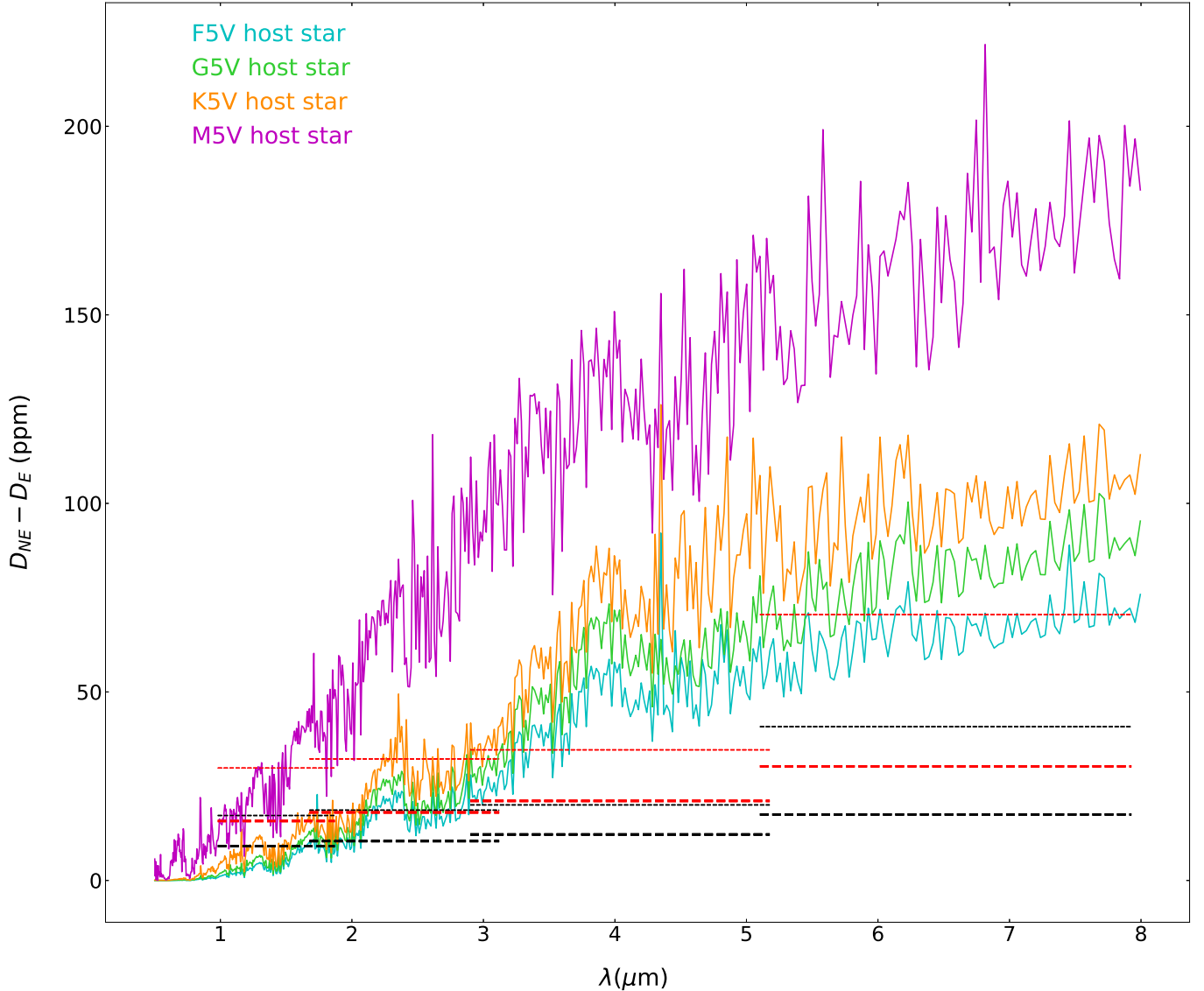}
\caption{Difference between the models of transit depth without and with thermal emission from the hot-Jupiters with $T_n=1600$ K, g=30 m/s$^2$, $R_P/R_*$ = $1.4R_J/R_\odot$ ($\sim$0.144), orbiting around stars of different spectral types. Transit depth without emission is independent of the host star spectral type. The 1-$\sigma$ noise-levels are shown in dashed lines from left to right for the JWST channels NIRSpec G140M, NIRSpec G235M, NIRSpec G395M, and MIRI LRS (slitless) respectively. The red and black dashed lines correspond to noise-levels for number of observed transits equal to 2 and 4 respectively. The thick and thin dashed lines correspond to noise-levels for host stars with J-band magnitude of 8 and 10 respectively.
\label{fig:spec}}
\end{figure}

The transit depth with no planetary thermal emission, $D_{NE}$, 
is absolutely independent of the stellar spectrum, as evident from 
Equation \ref{eq:radtran}. The factor $\frac{F_P}{F_*}$ solely depends on 
the physical and chemical properties of the planetary atmospheres and provides only the reduction in the stellar flux due to absorption. However, 
Equation \ref{eq:rademn} suggests that, when planetary emission is included,
the transit depth, $D_E$, for planets with the same $T_{n}$, becomes
dependent on the flux of the host star. Consequently, it depends on the
spectral types of the host stars. Figure \ref{fig:spec} displays the difference between $D_{NE}$ and $D_E$ for planets with $R_P/R_*$ = $1.4R_J/R_\odot$ ($\sim$0.144), g=30m/s$^2$  and $T_{n}=$1600K  orbiting stars of spectral types F5V, G5V, K5V, and M5V. Also, the 1-$\sigma$ noise levels of the JWST instrument modes NIRSpec G140M, NIRSpec G235M, NIRSpec G395M, and MIRI LRS (slitless) for number of observed transits equal to 2 and 4 and host stars with J-band magnitude of 8 and 10 are shown in this figure. The model spectra show  that
 the transit depths $D_E$ for stars of different spectral types 
differ significantly at wavelengths longer than 3 $\mu m$. The cooler the host stars are, the more significant the difference is between the models with respect to the noise levels, as evident from Equation \ref{eq:tdne2tde}.

\subsection{Average Night-Side Temperature and Planetary Size}

\begin{figure}[!ht]
\centering
\includegraphics[scale=0.43,angle=0]{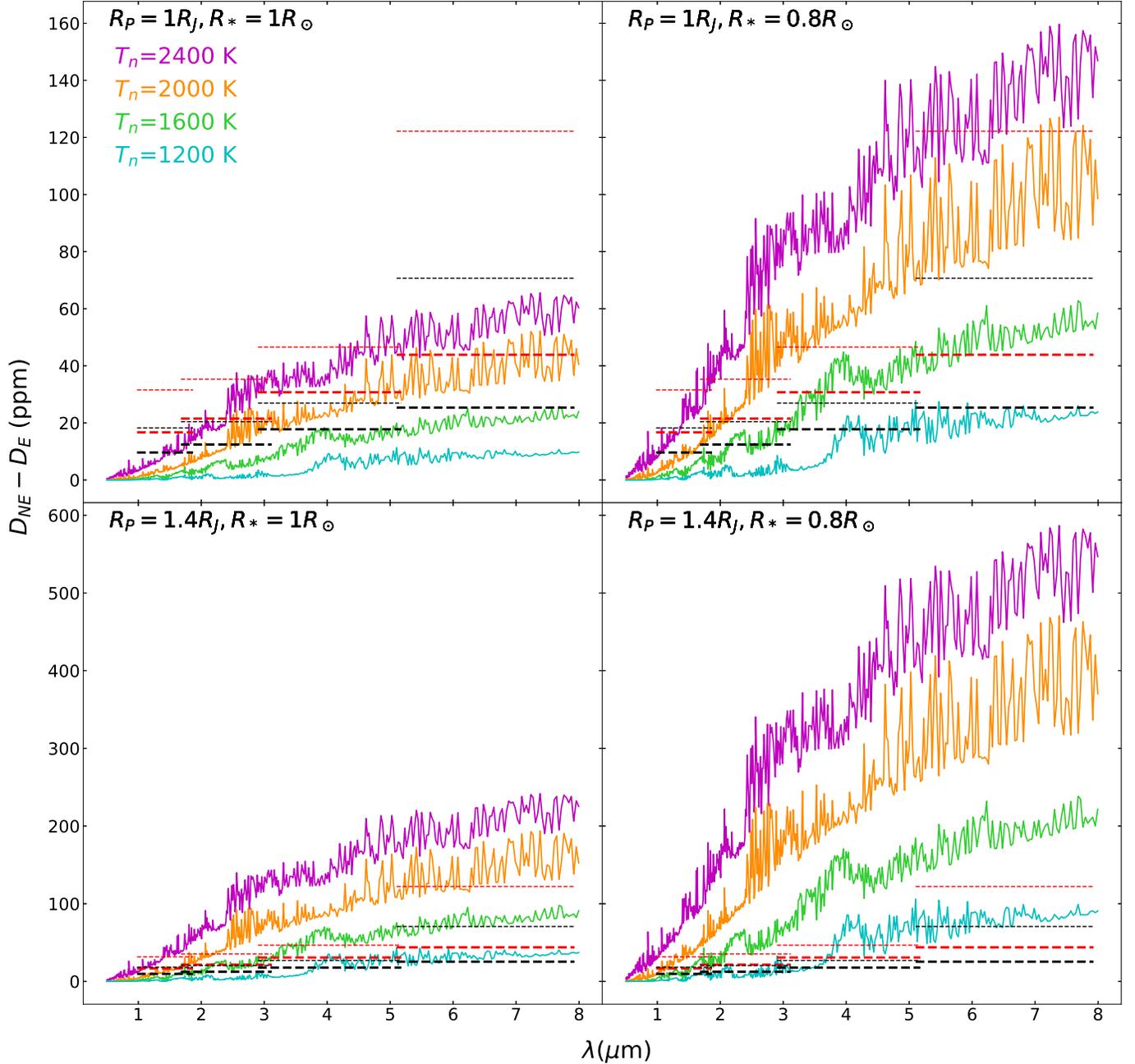}
\caption{Difference between the models of transit depth without and with planetary thermal emission for different values of $T_n$, $R_P$ and $R_*$, keeoing g fixed at 30m/s$^2$. The 1-$\sigma$ noise-levels are shown in dashed lines from left to right for the JWST channels NIRSpec G140M, NIRSpec G235M, NIRSpec G395M and MIRI LRS (slitless) respectively. The red and black dashed lines correspond to noise-levels for number of observed transits equal to 2 and 4 respectively. The thick and thin dashed lines correspond to noise-levels for host stars with J-band magnitude of 8 and 10 respectively.
\label{fig:teq_rprs}}
\end{figure}

In order to investigate the effect of $T_n$, we calculate the difference between $D_{NE}$ and $D_E$ at different values of $T_{n0}$ e.g., 1200K, 1600K, 2000K and 2400K.  Figure 13 of \cite{parmentier14} demonstrates that the Bond albedo of planets with high equilibrium temperature is extremely low ($<0.01$) for solar composition. Consequently, from Equation \ref{eq:tn} it follows that, $T_n\approx T_{n0}$. These values of $T_n$ correspond to atmospheric scale heights of 138 km, 186 km, 254 km and 387 km respectively. Figure \ref{fig:teq_rprs} shows
the difference between $D_{NE}$ and $D_E$ for these values of $T_n$ and for different values of $R_P$ and $R_*$ e.g., (i) $R_P=1R_J$, $R_*=1R_\odot$, (ii) $R_P=1R_J$, $R_*=0.8R_\odot$, (iii) $R_P=1.4R_J$, $R_*=1R_\odot$ and (iv) $R_P=1.4R_J$, $R_*=0.8R_\odot$. It is clear from the figure that with the increase in $T_n$, the difference between $D_{NE}$ and $D_E$ increases because the thermal re-emission from the planet increases. The factor $R_P/R_*$ also strongly dictates the significance of the difference with respect the noise-levels. This happens due to the fact that, with increasing $R_P/R_*$, the ratio of the thermal luminosity of the planet to the luminosity of the host star increases.
Obviously, for a fixed planetary radius, the difference in $D_{NE}$ and $D_E$ increases with the decrease in the size of the host star. 

Also, the 1-$\sigma$ noise levels of the JWST instruments NIRSpec G140M, NIRSpec G235M, NIRSpec G395M and MIRI LRS (slitless) for number of observed transit equal to 2 and 4 and host stars with J-band magnitude of 8 and 10 are shown in Figure \ref{fig:teq_rprs}. This helps us comprehend the significance of the difference between the models with respect to the noise levels for different numbers of observed transits and different host star J-band magnitudes. However, it can be safely ascertained that for higher values of $R_P/R_*$ (see bottom 2 panels of Figure \ref{fig:teq_rprs}) and for $T_n>1200$ K, the deviation of $D_E$ from the standard model of transmission spectra ($D_{NE}$) at wavelength beyond 2 $\mu m$ is so significant  that observed transit spectra can be misinterpreted by the standard model by 10-300 $\sigma$ (representing a change up to 0.5$\%$ in transit depth).

\subsection{Surface Gravity}

\begin{figure}[!ht]
\centering
\includegraphics[scale=0.4,angle=0]{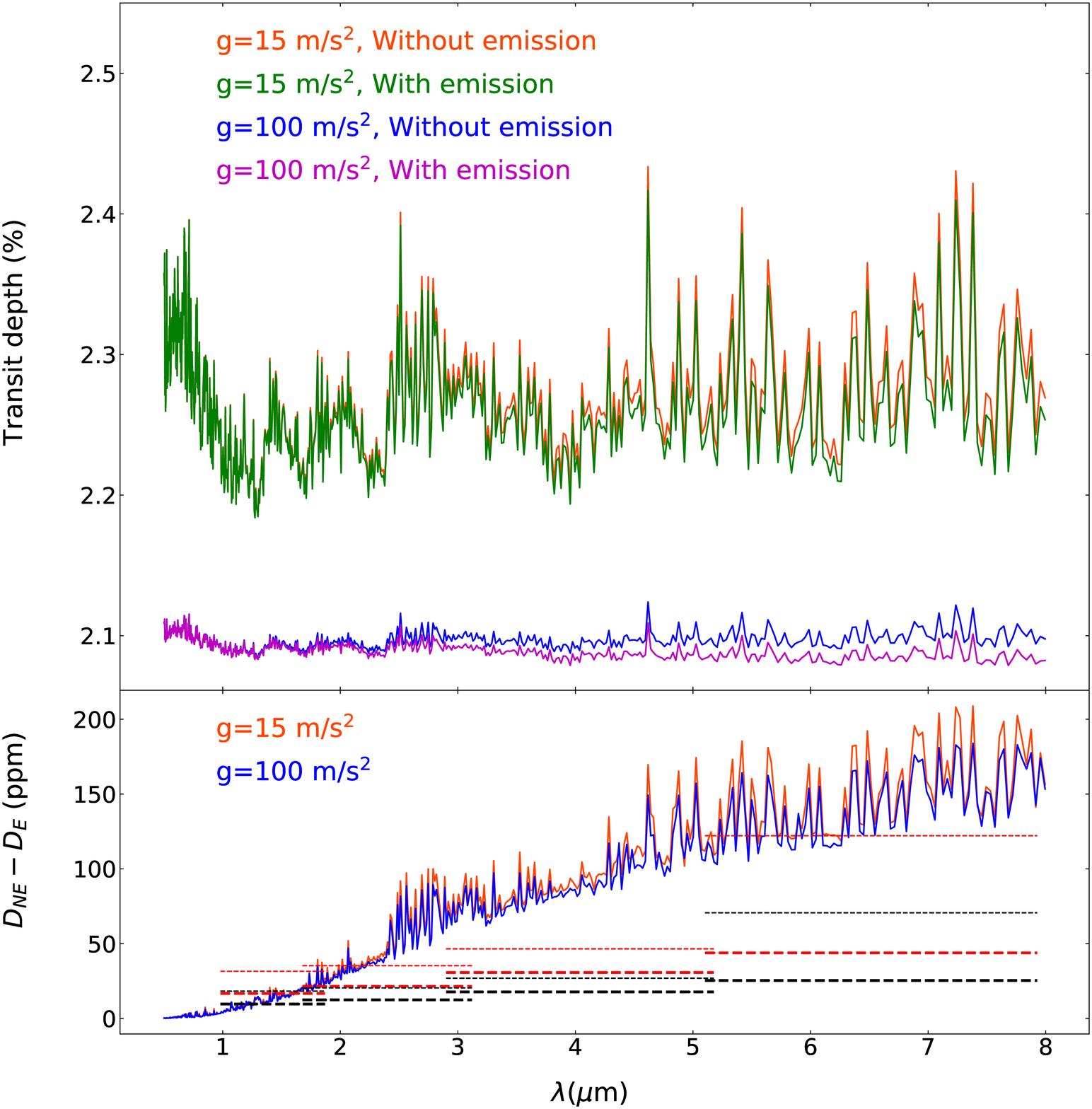}
\caption{Top - Models of transit depth without and with thermal emission for different values of $g$ and for $T_n$=2000K, $R_P$=1.4$R_J$ and $R_*$=1$R_\odot$. Bottom - Difference between the above models for each value g which shows no dependance on g. The 1-$\sigma$ noise-levels are shown in dashed lines from left to right for the JWST channels NIRSpec G140M, NIRSpec G235M, NIRSpec G395M and MIRI LRS (slitless) respectively. The red and black dashed lines correspond to noise-levels for number of observed transits equal to 2 and 4 respectively. The thick and thin dashed lines correspond to noise-levels for host stars with J-band magnitude of 8 and 10 respectively.
\label{fig:g}}
\end{figure}

We have calculated $D_{NE}$ and $D_E$ for g=15, 30, 60 and 100 m/s$^2$ as shown in Figure \ref{fig:g} for a fixed values of $T_n=$2000K, $R_P=1.4 R_J$ and $R_*=1 R_\odot$. These values of g correspond to atmospheric scale heights, estimated by using $T_n$=2000K, of 508 km, 254 km, 127 km and 76 km respectively. We find that with increasing g the transmission flux $F_P$ decreases. However, the value of g has almost no effect on the calculation of $F_{Th}$ and hence, the difference between $D_{NE}$ and $D_E$ is almost independent of g, the surface gravity of the planet.

\subsection{Atmospheric Clouds} \label{subsec:clouds}
\begin{figure}[!ht]
\centering
\includegraphics[scale=0.46,angle=0]{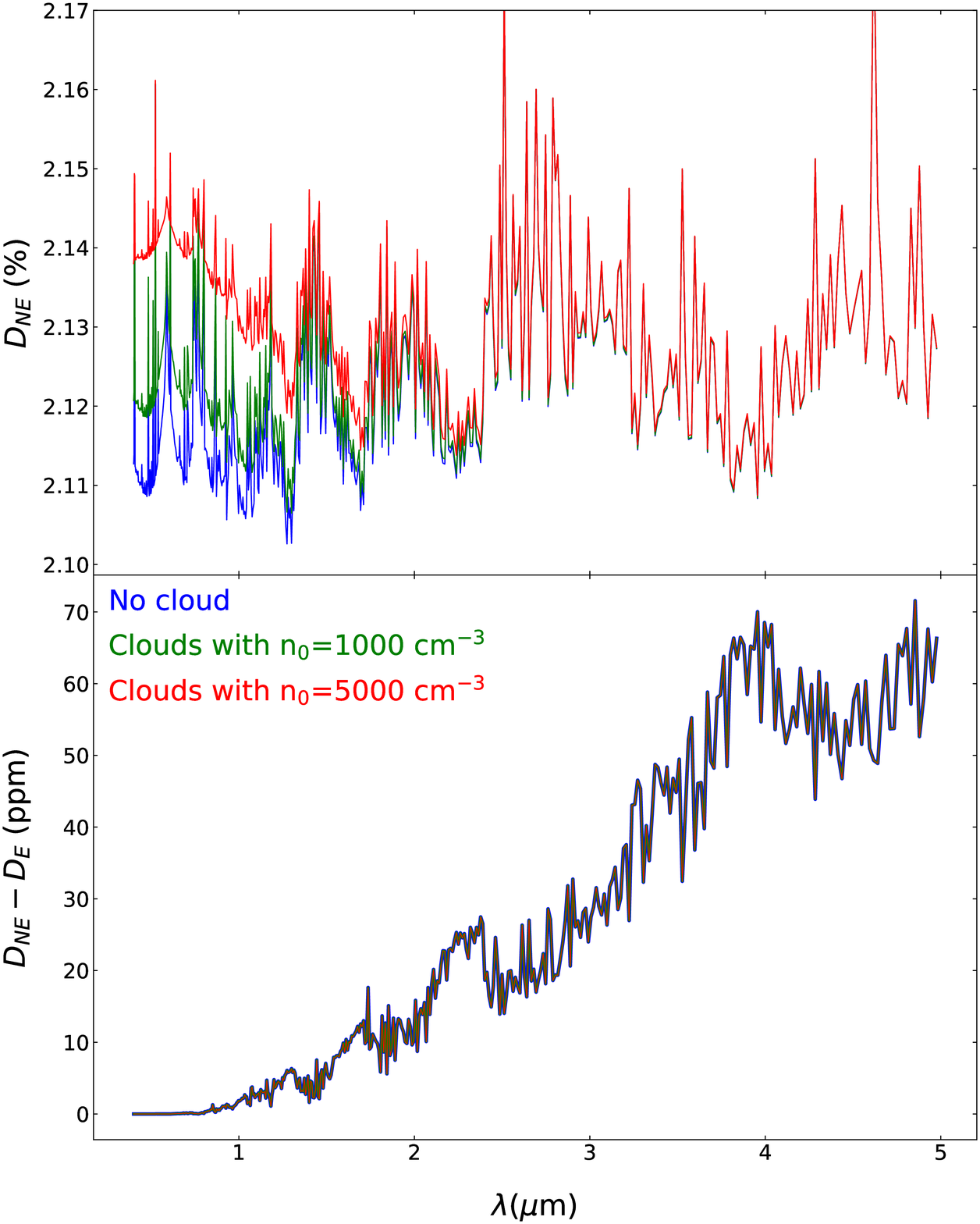}
\caption{Top - Models of transit depth with thermal emission with different cloud abundance as well as without any cloud for $T_n$=1600K, g=30 m/s$^2$, $R_P$=1.4$R_J$ and $R_*$=1$R_\odot$. Bottom - Difference between the models without and with thermal emission from the planets for the above cloud abundances as well as no cloud.
\label{fig:clouds}}
\end{figure}

Clouds and hazes are a ubiquitous feature in the planetary atmospheres. For hot exoplanets, silicates may condensate in the upper atmosphere. Gas giant planets with comparatively cooler night sides can have thick atmospheric clouds that may affect the spectra in the optical and near infrared wavelength region. However,  at higher day or night temperatures, clouds may either completely evaporate or may form a thin layer of haze in the uppermost atmosphere. As discussed in the previous subsection, the effect of thermal emission is significant only at wavelengths longer than 2$\mu$m and for a night-side temperature $T_n>1200$ K.  Therefore, even in the presence of a thin layer of haze, we don't expect the transmission spectra of planets having $T_n>1200$ K to be affected in the infra-red region where thermal re-emission is important.  Nevertheless, we have investigated the effects of the thin clouds on the transmission spectra of a planet with $T_n=1600$ K.  For this purpose, we have considered a simple model for thin haze in the uppermost atmosphere. The formalism is adopted from \cite{griffith98,saumon00}. We consider grains of amorphous Forsterite (Mg$_2$SO$_4$) of mean diameter 0.5 $\mu m$ as the dominant constituent of the cloud 
located within a thin region of the atmosphere bound by a base and a deck. Within this region, the sizes of the particles follow a log-normal distribution and the vertical density distribution of the cloud particles follows the relation 
$$n(P)=n_0\frac{P}{P_0}$$ 
where, n(P) is the number density of cloud particles at pressure level P, $P_0$ is the pressure at the base radius, and $n_0$ is a free parameter with the dimension of number density. The details of the model adopted can be found in \cite{sengupta20}. The deck and base of the haze are fixed at 0.1 Pa and 100 Pa pressure levels respectively. We use the Mie theory of scattering to calculate the wavelength-dependent scattering coefficients, extinction coefficients, and the phase functions at different pressure atmospheric depth \citep[][etc.]{vandehulst57,hansen74,fowler83,bohren83,sengupta09,sengupta20}. Figure \ref{fig:clouds} shows $D_{NE}$ (in top panel) and the difference between $D_{NE}$ and $D_E$ (bottom panel) for cloud models with $n_0=1000$ cm$^{-3}$ and $n_0=5000$ cm$^{-3}$. We compare the results with that of a cloud-free atmosphere.  Although the transit depth without thermal emission, $D_{NE}$,  can alter depending on the cloud structure and opacity, the difference between $D_{NE}$ and $D_E$ does not change much as the emission flux ($F_{Th}$) is not affected significantly in the infra-red region by the presence of cloud. Figure \ref{fig:clouds} shows that the difference between $D_{NE}$ and $D_E$ does not change at all with clouds. Hence, the clouds do not play an important role in determining the transit depth at the infra-red wavelength region of ultra-hot Jupiters.

\section{Conclusion} \label{sec:con}

We demonstrate that the effect of thermal re-emission from
the night side of hot Jupiters on the transit spectra can be significant at the infrared wavelength region if the equilibrium  temperature of the planet is higher than about 1200K and if the planet is large enough in size such that $R_P/R_* >$ 0.1.
The contribution of planetary thermal emission to the transit spectra can
significantly exceed the total noise budget (photon noise plus readout noise) of the
IR instruments on-board the upcoming  JWST that will perform  transit
spectroscopy.  Hence,
 a retrieval model that does not include planetary thermal emission would
overestimate the transit depth and thus can lead to
 a wrong interpretation of the planetary properties of the hot Jupiters.
Therefore, for a consistent and accurate interpretation of the observed
transit spectra, it is essential to include the diffused reflection
and transmission due to scattering in the optical and near infrared wavelength region and the thermal re-emission at the near and mid infrared region of hot gas giant planets.  Both need the solutions of the multi-scattering radiative transfer equations.

 \acknowledgments

 We thank the reviewer for several valuable comments and suggestions. 

\software{Exo\_Transmit \citep{kempton17}, Analytical$\;$model$\;$for$\;$irradiated$\;$atmosphere
 \citep{parmentier14,parmentier15}, Pandexo \citep{batalha17}}

\end{document}